\documentclass[prl,twocolumn,superscriptaddress,floats,showpacs]{revtex4}

\usepackage[dvips]{graphicx}
\usepackage{amsmath}
\usepackage{booktabs}
\usepackage{dcolumn}

\newcommand{\Ang}{$\text{\AA}$}
\newcommand{\Angrz}{$\text{\AA}^{-1}$}

\newcommand{\tc}{T$_\text{C}$}

\newcommand{\vQ}{$\vec{Q}$}

\newcommand{\bafeas}{BaFe$_2$As$_2$}
\newcommand{\bafecoasx}{Ba(Fe$_{1-x}$Co$_x$)$_2$As$_2$}
\newcommand{\bafecoass}{Ba(Fe$_{0.94}$Co$_{0.06}$)$_2$As$_2$}
\newcommand{\bakfeas}{Ba$_{0.6}$K$_{0.4}$Fe$_2$As$_2$}

\begin{document}

\advance\vsize by 2 cm

\title{Splitting of resonance excitations in nearly optimally doped Ba(Fe$_{0.94}$Co$_{0.06}$)$_2$As$_2$:
an inelastic neutron scattering study with polarization analysis}

\author{P. Steffens}
\email{steffens@ill.eu} \affiliation{Institut Laue Langevin, 6 Rue
Jules Horowitz BP 156, F-38042 Grenoble CEDEX 9, France}

\author{C. H. Lee}
\affiliation{National Institute of Advanced Industrial Science and Technology (AIST), Tsukuba, Ibaraki 305-8568, Japan}
\affiliation{Transformative Research Project on Iron Pnictides (TRIP), JST, Chiyoda, Tokyo 102-0075, Japan}

\author{N. Qureshi}
\affiliation{II. Physikalisches Institut, Universit\"{a}t zu K\"{o}ln, Z\"{u}lpicher Str.\ 77, D-50937 K\"{o}ln,
Germany}

\author{K. Kihou}
\affiliation{National Institute of Advanced Industrial Science and Technology (AIST), Tsukuba, Ibaraki 305-8568, Japan}
\affiliation{Transformative Research Project on Iron Pnictides (TRIP), JST, Chiyoda, Tokyo 102-0075, Japan}

\author{A. Iyo }
\affiliation{National Institute of Advanced Industrial Science and Technology (AIST), Tsukuba, Ibaraki 305-8568, Japan}
\affiliation{Transformative Research Project on Iron Pnictides (TRIP), JST, Chiyoda, Tokyo 102-0075, Japan}

\author{H. Eisaki}
\affiliation{National Institute of Advanced Industrial Science and Technology (AIST), Tsukuba, Ibaraki 305-8568, Japan}
\affiliation{Transformative Research Project on Iron Pnictides (TRIP), JST, Chiyoda, Tokyo 102-0075, Japan}

\author{M. Braden}
\email{braden@ph2.uni-koeln.de} \affiliation{II. Physikalisches Institut, Universit\"{a}t zu K\"{o}ln, Z\"{u}lpicher
Str.\ 77, D-50937 K\"{o}ln, Germany}

\date{\today}


\begin{abstract}

Magnetic excitations in Ba(Fe$_{0.94}$Co$_{0.06}$)$_2$As$_2$ are studied by polarized inelastic neutron scattering
(INS) above and below the superconducting transition. In the superconducting state we find clear evidence for two
resonance-like excitations. At a higher energy of about 8\ meV there is an isotropic resonance mode with weak
dispersion along the c-direction. In addition we find a lower excitation at 4 meV that appears only in the c-polarized
channel and whose intensity strongly varies with the $l$ component of the scattering vector. These resonance
excitations behave remarkably similar to the gap modes in the antiferromagnetic phase of the parent compound
BaFe$_2$As$_2$.

\end{abstract}

\maketitle

In various FeAs-based superconductors, INS studies have shown the existence of nearly antiferromagnetic excitations,
and in particular the resonance-like intensity enhancement upon entering the superconducting phase suggests a close
connection between superconductivity and magnetism \cite{hirshfeld}. Since the first observations of this resonant
feature in nearly optimally-doped \bakfeas\ \cite{nat-christ} and \bafecoasx\ (x=8\%) \cite{lumsden}, this signal has
been observed at different doping levels in the underdoped, optimally doped, and overdoped region
\cite{nat-christ,lumsden,chi,christianson,parshall,inosov09,lester10,wang10,pratt10,matan10,park10,li10,liu12nat,luo12,tucker12}
. Near optimum doping, in the paramagnetic phase and above \tc\ the magnetic excitations are nearly antiferromagnetic
spin fluctuations with anisotropic extension in longitudinal and transverse direction. These excitations can be
followed up to high energy \cite{lester10,li10,tucker12} where they strongly resemble those in the antiferromagnetic
parent compound \cite{lester10,matan10,liu12nat}. In the superconducting phase, spectral weight is suppressed at low
energy, and a resonant enhancement of the intensity is seen at an energy depending on the doping level (about 9\ meV
near optimum doping \cite{inosov09,lumsden}). Comparing different Fe-based systems the resonance energy seems to follow
a linear relation with $T_c$ \cite{park10}. In doped \bafeas\ the resonance mode appears to be quite broad and there is
sizeable dispersion along (0.5,0.5,$L$) \cite{wang10,pratt10} with the energy minima appearing at odd $L$.

In 5\% Ni-doped \bafeas, signs of significant magnetic anisotropy have been observed in polarized INS experiments
\cite{lipscombe}. It was concluded that the resonance mode exhibits a pure in-plane polarization at this doping level
while the out-of-plane excitations would exhibit a lower energy scale but no resonance mode. In contrast, at a
Ni-doping of 7.5\%, the magnetic excitations appear perfectly isotropic\cite{liu12}. A polarized neutron study of
FeTe$_{0.5}$Se$_{0.5}$ also finds more isotropic behavior with the resonance mode appearing in both channels at the
same energy, just the strength of the in-plane component is slightly larger\cite{babkevich}. Polarized INS has also
been used to study the gap of magnetic excitations in pure \bafeas\ that usually is attributed to single-ion
anisotropy, but in contrast to a simple easy-plane system it costs more energy to rotate the spin within the layer than
rotating it perpendicular to the layer\cite{qureshiBafeas}.

\begin{figure}
\begin{center}
\includegraphics*[width=.75\columnwidth]{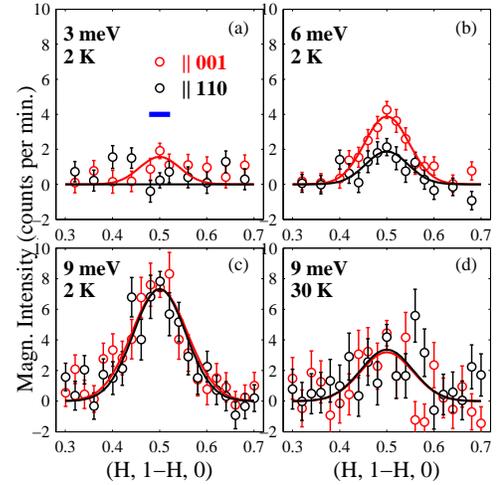}
\end{center}
\caption{(Color online) Constant-energy scans across Q=(0.5,0.5,0)
at different energy transfers and temperatures in the
superconducting and normal phase. Red: the scattered intensity due
to the component polarized along the c-axis of the crystal, black:
the component polarized along the diagonal of the ab-plane. The
horizontal bar in (a) is the expected width (fwhm) due to the
spectrometer's resolution. \label{figQscans}}
\end{figure}

Here, we present first results of polarized INS experiments on a 6\% \ Co-doped \bafeas\ crystal. In the
superconducting phase we find sizeable anisotropy in the magnetic response and demonstrate the existence of an
additional out-of-plane polarized resonance excitation appearing at lower energy. The spectra in the superconducting
phase remarkably resemble those taken in the antiferromagnetic state of the parent compound \cite{qureshiBafeas}.

\begin{figure}
\begin{center}
\includegraphics*[width=.9\columnwidth]{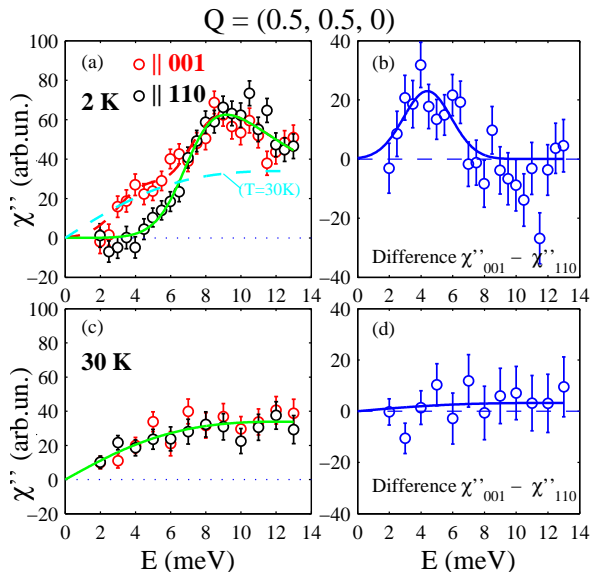}
\end{center}
\caption{(Color online) Energy dependence of the imaginary part of the generalized magnetic susceptibility at
Q=(0.5,0.5,0), for the out-of-plane and in-plane components, at T=2K (a,c) and T=30K (b,d). In (b) and (d) the
difference of the two components is shown. For comparison the isotropic susceptibility obtained at T=30K is shown in
part (a) as a dashed line. The raw count rates were corrected for the thermal factor and for the monitor effects as
described in the text, yielding a quantity proportional to the imaginary part of the magnetic susceptibility.
\label{figL0}}
\end{figure}

We have used a single crystal of \bafecoasx\ with Co-doping of x=6\% and with a mass of 2.85~g that was grown by the
FeAs-flux method. The sharp superconducting transition temperature of this sample has been determined as \tc=24~K by a
susceptibility measurement. Throughout the article, we use the tetragonal unit cell with a=b=3.95~\Ang\ and
c=12.98~\Ang. For the neutron scattering experiment, the crystal has been mounted in two different orientations, once
with [100] and [010] in the scattering plane, and once with [110] and [001].

We have first performed a diffraction study on the IN3 spectrometer to search for traces of magnetic ordering at the
(0.5,0.5,1) position, but did not find any intensity larger than 10$^{-4}$ times that of the fundamental (110) Bragg
peak. (In this configuration also the incommensurate peaks reported for Co-underdoped \bafeas\ \cite{pratt11} would
have been integrated by the resolution of the instrument.) Evidence for an onset of weak orthorhombic splitting has
been found by following the strong nuclear Bragg peak (020) as a function of temperature \cite{note020}.  We deduce
that our crystal exhibits an orthorhombic phase which gets suppressed by the onset of superconductivity similar to
previous reports \cite{nandi10}. The value of $T_c$, the observation of the structural distortion and its suppression
in the superconducting state, as well as the absence of antiferromagnetic ordering are in perfect agreement with
previous reports for 6\% \ Co-doped \bafeas \cite{pratt09,nandi10,pratt11}.

We have measured the magnetic excitations by polarized neutron scattering on the IN20 and IN14 thermal and cold triple
axis spectrometers at the Institut Laue-Langevin in Grenoble. The IN20 spectrometer was equipped with a Heusler
monochromator and analyzer and a graphite filter, with the final wave vector of the neutron fixed to 2.662~\Angrz,
while on IN14 we have used 1.55~\Angrz\ with a cooled Be-Filter. In both cases the sample was mounted in the Cryopad,
which insures a zero magnetic field environment of the sample. The flipping ratio, determined on a nuclear Bragg
reflection both in the normal and in the superconducting phase amounts to R=12, corresponding to a beam polarization of
85\% (and R$\sim$25 on IN14).

We have used a standard procedure to extract the magnetic signal from the spin-flip (SF) and non-spin-flip (NSF) cross
sections, with the neutron spin along the different directions parallel and perpendicular to the scattering vector \vQ,
as described for instance reference \onlinecite{qureshiBafeas}.

At all points, we have measured the three cross sections $\sigma^{x,y,z}_{SF}$, as well as $\sigma^x_{NSF}$. At
selected positions we have also measured $\sigma^{y,z}_{NSF}$ to check for consistency, which yielded good agreement.
Except a smooth decrease with scattering angle, the scattering in the $\sigma^x_{NSF}$ channel is featureless,
indicating the absence of sizeable nuclear contributions in the studied regions of \vQ ,$\omega$ space. The background
in the SF channels can be estimated by linear combination of the SF cross sections and showed no significant features
neither. It turns out that in all data sets the background can be well described by a single smooth weakly
scattering-angle dependent function, so we consider the extracted magnetic signal clean and free from contaminations.
In order to obtain the imaginary part of the magnetic susceptibility we have divided the scattering intensity by the
thermal population factor (Bose factor enhanced by one) and corrected for higher-order contributions in the incident
beam monitor.

\begin{figure}
\begin{center}
\includegraphics*[width=.9\columnwidth]{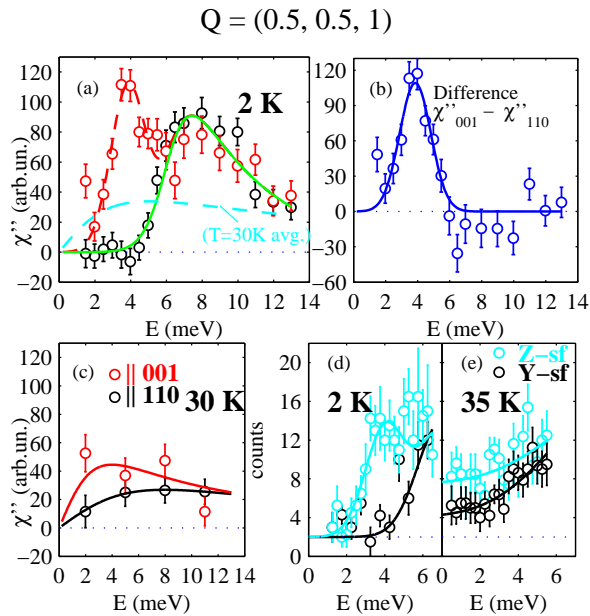}
\end{center}
\caption{(Color online) Energy dependence of the two components of magnetic susceptibility at \vQ =(0.5,0.5,1) at T=2\
K (a) and T=30\ K (c), and their difference at 2~K (b). The $c$-polarized component of the susceptibility exhibits a
sharp resonant mode in the superconducting phase. Panel (d) and (e) show low-energy raw data (not corrected for thermal
factor) of the Y- and Z-spin-flip channel collected in a separate measurement (IN14). The Z-sf channel contains mostly
magnetic excitations polarized along 001, and Y-sf along 110. Lines are guides to the eye, and the dashed line the
approximated background. \label{figL1}}
\end{figure}

In Figure \ref{figQscans} we show the results of constant-energy scans across \vQ =(0.5,0.5,0) in transverse direction,
where we superpose the signal stemming from magnetic excitations polarized perpendicular to the ab-plane and parallel
to it. The width of the signal is larger than the instrumental resolution. At an energy transfer of about 9~meV, we
observe the enhancement of intensity by the resonance in the superconducting state ($I_{2K}\approx 2.3\cdot I_{30K}$),
consistent with previous studies\cite{lumsden,li10}. Our polarized data show that the magnetic excitations at this
resonance energy are fully isotropic in spin space. At lower energy transfer, however, the component along the c-axis
is significantly stronger. At 6~meV, the anisotropy factor is about 2, and at 3~meV, where most intensity is suppressed
in the in-plane channel, still a very small component along $c$ remains.

The energy scans shown in Fig.\ \ref{figL0} confirm this behavior. At 2~K, the magnetic response consists of an
isotropic component, giving rise to the resonance peak, and of an additional feature polarized along $c$, which appears
at half the energy of the resonance. The difference of the two components shown in Fig. 2(b) clearly illustrates the
isotropic character of the high-energy mode and the c-polarized nature of the lower signal. In the normal state
(T=30K), the magnetic response is rather smooth and can be well described by a single relaxor formula,
$\chi''(\omega)=\frac{\chi'\Gamma}{\omega^2+\Gamma^2}$, with $\Gamma$=12$\pm$2~meV. Above T$_c$ we find no significant
magnetic anisotropy at Q=(0.5,0.5,0).

We have also performed measurements at $L$=1, see Fig.\ \ref{figL1}, using the [110]/[001] scattering geometry. The
intensities cannot directly be compared with those in the other geometry, as the experimental resolution changes the
integration of the magnetic signal. Due to the large width of the signal in \vQ -space, though, the effect on the
intensity is not very large, and checking identical \vQ -positions shows that the count rate in the second orientation
(110-001) is only slightly reduced. At the energy of the resonance signal at 8\ meV, the count rate is nevertheless
found to be $\sim$30\% higher at $L$=1 indicating sizeable dependence of the resonance intensity on $L$. Previous
studies have found a stronger dependence for 4\% Co-doping\cite{christianson}, and almost none (besides the magnetic
form factor) for 8\% Co \cite{lumsden,parshall}. Most remarkably, the low-energy resonance feature is again observed in
the $c$ component of the susceptibility. For $L$=1, this low energy mode exhibits an even stronger signal than the
$c$-axis polarized component of the high-energy resonance mode. The additional sharp peak in the $\chi ''$ component
parallel $c$ possesses thus a pronounced dependence on $L$. At $L$=1, there is a small anisotropy also above \tc\
(Fig.\ \ref{figL1}c,e). A focus on the low energy part of the spectrum using the IN14 spectrometer (Fig.\
\ref{figL1}d,e) proves, however, that the normal state response is smooth, and that the low-energy signal is
intrinsically linked to the superconducting state.

The polarized INS experiment shows that magnetic excitations appearing in the superconducting state of \bafecoass \
possess an intrinsic structure and consist of two distinct components: the high-energy resonance which is magnetically
isotropic  and an additional low-energy component that is entirely polarized in the $c$ direction. This low-energy
uniaxial mode has three-dimensional character, as shown by its strong $L$ dependence, while the $L$ dependence of the
resonance is weaker, underlining a more two-dimensional character. The $c$-polarized excitation at low energy points to
a fundamental property of the superconducting state near optimum doping.

\begin{figure}
\begin{center}
\includegraphics*[width=.85\columnwidth]{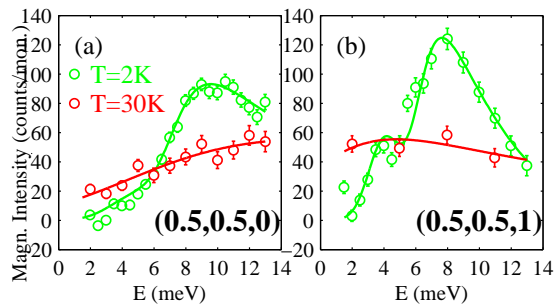}
\end{center}
\caption{(Color online) Total magnetic signal, obtained by the sum
of all spin-flip channels (SF$_X$+SF$_Y$+SF$_Z$)/2 (background
subtracted), above and below \tc\ and at $L$=0 (a) and $L$=1 (b).
Lines are guides to the eye, as in Figs.\ \ref{figL0} and
\ref{figL1}, counts are normalized to monitor which corresponds to
approx. 6.5 min.\ counting time. Note that the data in (a) and
(b) have been taken in different orientations of the sample.
\label{figsum}}
\end{figure}

By summing up the magnetic contributions, we obtain the curves shown in Fig.\ \ref{figsum} which correspond to the
signal expected in an unpolarized INS experiment. Indeed they closely resemble the published unpolarized data near
optimum doping\cite{lumsden,inosov09,li10,matan10}. Because the low-energy component contributes only to one channel,
it has a relatively small weight in the summed response, rendering it hard to detect without polarization analysis.
However, close inspection of published unpolarized data on Co and Ni doped \bafeas\ strongly suggests that this signal
is present there as well but has been overlooked due to the superposition with the isotropic response
\cite{inosov09,matan10,park10}. The dispersion of the resonance excitation has been so far determined by fitting the
total magnetic response in the superconducting phase (or the difference between superconducting and normal-conducting
phases) by a single peak \cite{wang10,pratt10}. Due to the strong variation of the low-energy signal with $L$ this
procedure partially hides the overlooked additional intensity in a more pronounced dispersion of the resonance energy.
The analysis of our polarized data for the high-energy resonance yields a weak dispersion (8.4meV at $L$=0 and 7.8meV
at $L$=1) and there is no significant dispersion of the low-energy signal.

We also may compare the polarized INS result to the Ni-doped \bafeas\ superconductors, for which polarization analysis
experiments have recently been performed. Near optimum Ni-doping\cite{lipscombe}, a pronounced anisotropy has been
found, but the interpretation disagrees with our analysis. Lipscombe et al. conclude that the high-energy resonance
mode appears only in the in-plane channel while the out-of-plane susceptibility would be feature-less in this energy
range an exhibit a lower energy scale\cite{lipscombe}. The much better statistics of our data for \bafecoass\ clearly
disproves such an interpretation in our case. We show that the high-energy resonance mode exists in both channels and
that the additional low-energy resonance mode appears in the out-of-plane channel. We think this interpretation is also
consistent with the statistics of the experiment on BaFe$_{1.9}$Ni$_{0.1}$As$_2$ \cite{lipscombe}, but an experiment
with better statistics is highly desirable. In the Ni overdoped compound the magnetic response has been found to be
completely isotropic\cite{liu12} excluding the possibility of a similar low-energy mode polarized along $c$. The fact
that the anisotropy in the magnetic excitations occurs most strongly around optimal doping may suggest that it
stabilizes the superconducting state.

A paramagnetic itinerant system may exhibit a strongly anisotropic susceptibility when approaching a magnetic
transition, as it has been observed in pure Sr$_2$RuO$_4$ \cite{srruo1} which is close to a spin-density wave (SDW)
ordering driven by Fermi surface nesting \cite{srruo2}. However, in the ruthenate, the channel which condenses into the
SDW state exhibits the larger amplitude and the lower characteristic energies, while here we find the opposite. In
\bafecoass, the out-of-plane polarization, which does not correspond to the order in the parent compound \cite{huang},
exhibits the lower resonance energy. Furthermore, Sr$_2$RuO$_4$ exhibits the anisotropy in its normal state, whereas
the anisotropy in \bafecoass\ emerges with the opening of the superconducting gap. However, the magnetic response of
superconducting \bafecoass\ is remarkably similar to that of undoped antiferromagnetic \bafeas. It has been outlined by
several groups that the high-energy magnetic response in the superconducting phase in the energy range of 20 to 200\
meV is very similar to that in the parent compounds in or above its N\'eel state
\cite{lester10,matan10,liu12nat,matanbafeas,harriger}. It was argued that the main difference consists in the fact that
modes depart from the resonance excitation in the superconductor while they start at the gap modes in the
antiferromagnetic phase in \bafeas . Comparing the polarized study of the gap modes in \bafeas \cite{qureshiBafeas}
with the results presented here reveals even more similarity. In both cases one finds a clear splitting of in-plane and
out-of-plane spin gaps with the c-polarized response appearing at lower energy. (Since magnetic moments are aligned in
the plane in \bafeas, the observed anisotropy is opposite to the expectation of a simple local-moment picture for the
pure compound.) 
We state that the opening of either a superconducting or of a SDW gap results in quite
similar magnetic spectra in spite of the different nature of these phases. The energy scale of the spin gaps induced is
however much smaller in the case of the superconducting compound.

The anisotropy in magnetic response indicates the presence of significant spin-orbit coupling. Orbital effects and the
possible existence of nematic order have been proposed to explain the in-plane anisotropy of the physical properties in
particular of the pure materials\cite{kruger,lee,lv}. The difference between in-plane and out-of-plane magnetic
excitations in \bafecoass\ also results from orbital degrees of freedom, and the smaller but qualitatively similar
effect compared to that in pure \bafeas\ suggests similar orbital features to be involved. The isotropic response in Ni
overdoped samples indicates that such orbital effects are strongest in the vicinity of the orthorhombic
antiferromagnetic phase where the highest superconducting $T_c$'s are observed.

In conclusion, the polarization analysis of magnetic excitations in \bafecoass\ has shown that the resonance
excitations in the superconducting phase are more complex than previously assumed. A broad signal at higher energy
exhibits no sizeable magnetic anisotropy in spin space, neither at $L$=0 nor at $L$=1. However, there is an additional
low-energy component which is polarized along the c-direction of the lattice, and whose scattering strength depends on
the $L$ value: it is weak for $L$=0 and very strong for $L$=1. Besides a lower energy-scale, the two resonance
excitations strongly resemble the low-energy response in pure antiferromagnetic \bafeas\ which is governed by two
different spin gaps for the two transverse magnetic polarizations. The magnetic excitations in undoped and in optimally
doped \bafeas\ appear thus much more similar than what one might expect for superconducting and antiferromagnetic
phases.

\paragraph*{Acknowledgments.} This work was supported by the Deutsche Forschungsgemeinschaft through SFB 608, and by a
Grant-in-Aid for Scientific Research B (No. 24340090) from the Japan Society for the Promotion of Science.


\begin{thebibliography}{}
\bibitem{note020} The (020) intensity slightly increases upon cooling due to the Debye-Waller factor but below about 40K
the intensity increase is drastically enhanced till $T_c$ where the signal drops down almost to values expected from
the normal intensity increase, resulting in a clear maximum at $T_c$. The width of the (020) peak shows no visible
effect. With the strong Bragg peak, we sense the extinction conditions in the large crystal. The orthorhombic phase
implies a domain structure and thereby reduces the mosaic quality and the extinction of the crystal. This results in an
intensity increase of strong Bragg peaks.
\end{thebibliography}
\end{document}